\documentclass[aps,prl,twocolumn,float,showpacs]{revtex4}
\usepackage{amsmath,bm,epsfig}

\usepackage{amssymb}
\usepackage{amssymb}
\usepackage{amscd}

\input epsf

\def\a{\alpha}
\def\b{\beta}
\def\g{\gamma}

\def\o{\omega} \def\O{\Omega}

\def\k{{\textbf{k}}} 

\def\be{\begin{equation}}
\def\ee{\end{equation}}
\def\bea {\begin{eqnarray}}
\def\eea {\end{eqnarray}}

\def\RA {\ \Rightarrow\ }

\begin{document}

\title{Effect of non-zero constant vorticity on the nonlinear resonances\\ of capillary water waves}%
\author{Adrian Constantin$^{\dag}$ and Elena Kartashova}%
\email{adrian.constantin@univie.ac.at, lena@risc.uni-linz.ac.at}
\affiliation{$^\dag$ Faculty of Mathematics, University of Vienna, 1090 Vienna, Austria\\
$^*$ RISC, J. Kepler University, Linz 4040, Austria}

\begin{abstract}
The influence of an underlying current on 3-wave interactions of
capillary water waves is studied. The fact that in irrotational flow
resonant 3-wave interactions are not possible can be invalidated by
the presence of an underlying current of constant non-zero vorticity.
We show that: 1) wave trains in flows with constant non-zero vorticity 
are possible only for two-dimensional flows; 2) only positive constant vorticities can trigger the
appearance of three-wave resonances; 3) the number of positive
constant vorticities which do trigger a resonance is countable; 4) the
magnitude of a positive constant vorticity triggering a resonance can not be too small.
\end{abstract}

\pacs{47.15.ki, 47.35.Pq, 47.10.A-}


\maketitle

\noindent \textbf{1. Introduction.} In this Letter we investigate
the effect of a current on the dynamics of nonlinear capillary water
waves, a problem of fundamental importance due to the ubiquity of
currents at sea \cite{Io}. The most common force for creating water
waves is the wind and wind generated capillary waves play a
prominent role in the development of waves on water surfaces that
are flat in the absence of wind. Indeed, capillary waves generate
surface roughness allowing a better grip of the wind. This leads to
the subsequent development of capillary-gravity and gravity waves as
with increasing wave amplitude gravity becomes the dominant
restoring force replacing surface tension \cite{Li, PS, S}. In
coastal navigation the important question arises whether the
presence of an underlying current can be detected by investigating
solely phenomena at the water's surface. Vorticity is adequate for
the specification of a current. A uniform current is described by
zero vorticity (irrotational flow), while the simplest example of a
non-uniform current is that of tidal flows, which can be
realistically modeled as two-dimensional flows with constant
non-zero vorticity, with the sign of the vorticity distinguishing
between ebb/tide \cite{DP}. Notice that in linear systems waves of
different frequencies do not interact due to the superposition
principle, while in a nonlinear system the lowest-order nonlinear
effect (with respect to an expansion in the wave amplitudes) is the
resonant interaction of three waves of different wave-vectors and
frequencies. Resonant interactions can profoundly affect the
evolution of waves by making significant energy transfer possible
among the dominant wave trains, accounting thus for wave patterns
that are higher and steeper than linear wave theory would predict
and providing insight into the effects of weak turbulence
\cite{PZ00}. It is known  that capillary waves in irrotational flow
do not exhibit exact 3-wave resonance \cite{K98}. We will show that
the presence of an underlying current of constant vorticity can
lead to the excitation of nonlinear resonances, but
only for special vorticities.\\

\noindent \textbf{2. Dispersion function.} Our
purpose is to show that currents strongly affect the dynamics already
at the level of capillary waves. To emphasize the wave-current interactions we
consider a setting with a flat bed in order to minimize the effect
of the shape of the shoreline and of the water bed on such flows. In
this context we investigate whether or not three-wave resonances are
possible among capillary waves. The dispersion relation for exact
solutions to the governing equations for capillary-gravity water waves
propagating at the free surface of
water of constant density $\rho=1$ with a flat bed and in a flow with constant vorticity $\Omega$
is
 \bea\label{dis}
 \omega & = & \frac{\Omega}{2}\,\tanh(kd)\nonumber \\
 & + & \frac{1}{2}
\sqrt{\Omega^2\,\tanh^2(kd) + 4 (kg+k^3\sigma) \,\tanh(kd)},\eea
where $d>0$ is the average water depth, $g$ is the gravitational
constant of acceleration and $\sigma>0$ is the coefficient of
surface tension, and the notation  $k$ is used for the modulus of the wave
vector, $k=|\k|$. The case of capillary waves corresponds to $g
\to 0$ and that of gravity waves to $\sigma \to 0$ cf. \cite{W0, W, CS}. Moreover,
capillary waves being short waves, it is appropriate
to take the short wave limit $kd \to \infty$ of the above dispersion
relation; as $\tanh(kd) \to 1$ we obtain the dispersion relation for
small-amplitude capillary waves as 
\be \label{DispCap}
\omega = \frac{\Omega}{2}  + \frac{1}{2}
\sqrt{\Omega^2 + 4 k^3\sigma }. \ee
In the case of irrotational flows (with zero vorticity), the dispersion relation reads \cite{W} 
\be \label{disp-ir-cap}\o^2=\sigma k^3.\ee%
As it was shown in \cite{K98}, exact three-wave resonances are
prohibited for (\ref{disp-ir-cap}) by purely kinematic considerations.
Below we show that in the case of non-zero constant vorticity,
resonances are  possible but only for a countable number of positive
vorticities.\\

\noindent \textbf{3. Magnitudes of vorticity.} First, we show that
the resonance conditions
 \be
\label{3resKin}\omega(\k_1)+\omega(\k_2)=\omega(\k_3), \quad
\k_1+\k_2= \k_3 \ee 
do have solutions for the dispersion function (\ref{DispCap}). Let
us introduce the variable $ \zeta=\Omega / 2 \sqrt{\sigma}$ and
rewrite (\ref{3resKin}) as
$$\sqrt{\zeta^2+k_1^3}+\sqrt{\zeta^2+k_2^3}=\sqrt{\zeta^2+k_3^3} -
\zeta, \quad \k_1+\k_2= \k_3 .$$

Our primary goal is investigating whether for a given current
strength $\Omega$ three-wave resonances are possible for wave
vectors with integer coordinates. Since capillary wave trains
propagating in flows of constant vorticity
are only possible if the flow is two-dimensional (see Section 5), only  the case of one-dimensional
wave-vectors, $\dim{\k}=1,$ is relevant. Now $\k=k$ and (\ref{3resKin})
reads
\bea
\sqrt{\zeta^2+k_1^3}+\sqrt{\zeta^2+k_2^3}=\sqrt{\zeta^2+k_3^3} - \zeta,\label{e-1} \\
k_1+k_2=k_3 ,\label{e-2} \eea
and we will study the magnitudes of vorticities corresponding  to
integer wave-vectors $k_1,\,k_2,\,k_3
> 0$. To investigate (\ref{e-1}),(\ref{e-2}) it is appropriate to view it
as an equation in the unknown $\zeta \neq 0$, with $k_1,\,k_2,\,k_3 > 0$
given integers. Let us first simplify notation by denoting
\begin{equation}\label{d}
k_1^3=\alpha,\quad k_2^3=\beta,\quad k_3^3=\gamma,\quad
E=\alpha+\beta-\gamma.
\end{equation}
Since irrespective of the sign of $\zeta$ both sides of (\ref{e-1})
are positive, squaring them we see that (\ref{e-1}) is equivalent to
\begin{equation}\label{e2}
E + 2\zeta
\sqrt{\zeta^2+\gamma}=-2\sqrt{(\zeta^2+\alpha)(\zeta^2+\beta)}.
\end{equation}
Once we obtain a formula for $\zeta^2$, equation (\ref{e2}) yields
\begin{equation}\label{y}
\zeta=-\,\frac{E+2\sqrt{(\zeta^2+\alpha)(\zeta^2+\beta)}}{2\sqrt{\zeta^2+\gamma}}.
\end{equation}
But (\ref{e2}) squared yields
$$E^2 -4\zeta^2E-4\alpha\beta=-4E\zeta\sqrt{\zeta^2+\gamma},$$
which, squared again, leads to
\bea (E^2-4\alpha\beta)^2 = \nonumber\\
 8E\,\zeta\,(2E\gamma+E^2-4\alpha\beta)=
8E\,\zeta^2\,[(\alpha-\beta)^2-\gamma^2],\nonumber \eea%
 with the last equality a
simple consequence of the definition of $E$. Thus
\begin{equation}\label{y2}
\zeta^2=\frac{(E^2-4\alpha\beta)^2}{8\,(\alpha+\beta-\gamma)[(\alpha-\beta)^2-\gamma^2]}.
\end{equation}
Obviously, (\ref{y2}) can be regarded as a 2-parameter series of
solutions for (\ref{e-1})-(\ref{e-2}). Indeed, for any wave-vectors $k_1,
k_2$ there are at most two vorticities $\Omega$ generating a
three-wave resonance of the form (\ref{e-1})-(\ref{e-2}). For instance,
$k_1= k_2=1$ yield $k_3=2, \ \a=\b=1, \ \g=8,\ E=-6, \ \zeta^2= 1/3$
and the corresponding vorticity $\Omega$ has to satisfy
$\Omega^2=4\sigma/3.$ Now we have to gain more understanding about
the possible sign of the vorticity. To formulate our theorem on the sign of vorticity,
let us first introduce a new variable%
 \be \label{width} \delta=
\sqrt{k_3^3}-  \sqrt{k_1^3} - \sqrt{k_2^3} . \ee %

\textbf{Theorem 1} (on the sign on vorticity). {\it Three
wave-vectors $k_1,\,k_2,\,k_3 \ge 1$ are
 solutions of (\ref{e-1}),(\ref{e-2}) if and only if $\delta>0$, in which case
 the vorticity $\Omega=2\sqrt{\sigma}\zeta$ is positive, with $\zeta$ given
 by (\ref{y}).} \smallskip

\textbf{Proof.} Two observations can be made immediately. Firstly, as
it was shown in \cite{K98}, the equality $\delta  = 0$ can be
transformed to the particular cubic case of Fermat's Last Theorem and
therefore has no integer solutions.  Secondly, condition (\ref{e-2}) yields%
\begin{equation}\label{c1}
k_3^3 > k_1^3+k_2^3
\end{equation}%
and therefore $E< 0$.

We first claim that if $k_1,\,k_2,\,k_3 \ge 1$ are solutions of (\ref{e-1}),(\ref{e-2}),
then $\delta >0$. Indeed, notice that the inequality
$$3 k_1^2k_2 + 3 k_1k_2^2 = 3 k_1k_2 ( k_1 + k_2 ) \ge 6 k_1k_2 \sqrt{k_1k_2}$$
yields $3 k_1^2k_2 + 3 k_1k_2^2 > 2 \sqrt{ k_1^3k_2^3 }$. If (\ref{e-2}) holds, adding $k_1^3+k_2^3$
to both sides of the above inequality, we get
$$k_3^3 \ge \Big(\sqrt{k_1^3}+
\sqrt{k_2^3}\Big)^2,$$
that is, $\delta>0$.

Let us now prove that for $\delta  > 0$ we always have a solution
$\zeta>0$ and no solution $\zeta<0$. The statement about the
positive solution follows at once by noticing that the function
$\zeta \mapsto \sqrt{\zeta^2+\alpha}
+\sqrt{\zeta^2+\beta}-\sqrt{\zeta^2+\gamma} +\zeta$ has limit
$+\infty$ for $y \to \infty$, being strictly increasing on
$(0,\infty)$ since its derivative satisfies
$$\frac{\zeta}{\sqrt{\zeta^2+\alpha}}+
\frac{\zeta}{\sqrt{\zeta^2+\beta}}-\frac{\zeta}{\sqrt{\zeta^2+\gamma}}+1
> \frac{\zeta}{\sqrt{\zeta^2+\alpha}} +1>1,$$
for $\zeta >0,$ as $\gamma>\beta$ in view of (\ref{c1}). Therefore
this function has a zero on $(0,\infty)$ if and only if its value at
$\zeta=0$ is strictly negative, that is, if and only if $\delta  >
0.$ Now  notice that we know by (\ref{y}) and (\ref{y2}) that if a
solution exists, it is unique. This concludes the proof.
 \smallskip

\textbf{Corollary 1} {\it The set of all resonant triads covered by
(\ref{e-1}),(\ref{e-2}), is generated by a countable number
of positive vorticities.}\smallskip

More precisely, given the wave-vectors $k_1,\,k_2 \ge 1$ we define $k_3$
via (\ref{e-2}), and compute
$$\begin{array}{l}
E=-3k_1k_2\,(k_1+k_2),\\[0.1cm]
E^2-4\alpha\beta = k_1^2k_2^2\,(9k_1^2+9k_2^2+14k_1k_2),
\end{array}$$
finding also that $\gamma^2-(\alpha-\beta)^2$ equals
$$k_1k_2\,(6k_1^4+15k_1^3k_2 +22k_1^2k_2^2
+15 k_1k_2^3+6k_2^4).$$ Knowing that $\zeta>0$, we infer from
(\ref{y2}) that a three-wave resonance occurs only if
$\Omega$ is
given explicitly by%
\be\label{om} \frac{\sqrt{\sigma}\,k_1^{3/2}k_2^{3/2}\,(9k_1^2+9k_2^2+14k_1k_2)}{\sqrt{6(k_1+
k_2)(6k_1^4+15k_1^3k_2 +22k_1^2k_2^2 +15 k_1k_2^3+6k_2^4)}}.\ee
For $k_1 \ge k_2$ we see that the denominator is bounded from
above by $16\sqrt{3}\,k_1^{5/2}$ and from below by $16\sqrt{3}\,k_2^{5/2}$. Since
$$9k_1^2+9k_2^2+14k_1k_2 = 9(k_1-k_2)^2+32k_1k_2 \ge 32k_1k_2,$$
a lower bound for the numerator is $32\,\sqrt{\sigma}\,k_1^{5/2}k_2^{5/2}$, with the
evident upper bound $32\,\sqrt{\sigma}\,k_1^{7/2}k_2^{3/2}$. Therefore
$$2\,\sqrt{\frac{\sigma}{3}} \,k_1^{5/2}\,\frac{k_1}{k_2} \ge \Omega \ge  \smallskip
2\,\sqrt{\frac{\sigma}{3}} \,k_2^{5/2}.$$
\smallskip

\textbf{Corollary 2} {\it Three-wave resonant interactions do not
occur in flows with sufficiently small constant vorticity, the
minimal magnitude of a resonance generating vorticity being
$\O_{min}=2\sqrt{\sigma}/\sqrt{3}.$} \\

\noindent \textbf{4. Structure of resonances.} In the previous
section we have shown that for any two wave-vectors $k_1, k_2$, a
corresponding resonant triad $(k_1, k_2, k_1+k_2)$ will be generated
by the vorticity given by (\ref{om}). The inverse problem --- to
compute all resonant triads generated by a given vorticity --- is
much more complicated. Already for one fixed wave-vector, this
problem is equivalent to finding rational points on an elliptic
curve. On the other hand, it is relatively easy to compute
numerically, in a some finite spectral domain, the set of all
resonant triads and the corresponding magnitudes of the vorticity.
 Another reasonable
question is whether different resonant triads can be generated by
almost the same magnitude of vorticity, with some accuracy
$\epsilon$ because these type of resonances might enrich the cluster
structure.

The structure of resonance clusters
has been investigated numerically, in the spectral domain $k_1, k_2
\le 100.$ All exact solutions for $\zeta^2$, that is, solutions with
$\epsilon =0$ have one of two forms:

1. Each resonance triad of the form $(k_1, k_1, 2k_1)$ with
$k_1=1,2,..., 100$ can be generated by only one vorticity; two
different triads of this form, $(k_1, k_1, 2k_1)$ and  $(k_2, k_2,
2k_2)$ with $k_1 \neq k_2$  can only be generated by  two
different vorticities.

2. Each resonant triad of the general form $(k_1, k_2, k_1+k_2)$
with $k_1 \neq k_2$ has the corresponding symmetrical resonant triad
$(k_2, k_1, k_1+k_2)$. Both symmetrical triads are generated by the
same vorticity. This fact can be observed immediately from
the form of (\ref{om}) which is invariant under the
transformation $[k_1 \mapsto k_2, \  \ k_2 \mapsto k_1].$

It appears that a given vorticity can generate not more than two
exact resonant triads. The situation does not change substantially
when $\epsilon \neq 0$: again, not more than two resonant triads appear to
be generated by almost the same magnitude
of vorticity. For example,%
\bea \label{Delta}%
\epsilon &=& 10^{-1} \RA (1,9,10), \quad \ \ \zeta^2= 4.824;\nonumber \\
\epsilon &=& 10^{-1} \RA (2,3,5), \quad \ \ \ \
\zeta^2=4.764;\nonumber\\
\epsilon &=& 10^{-2} \RA (2,14,16), \quad \ \zeta^2= 29.5622;\nonumber \\
\epsilon &=& 10^{-2} \RA (4,5,9), \quad \ \ \ \ \zeta^2= 29.5612;\nonumber \\
\epsilon &=& 10^{-4} \RA (2,88,90), \quad  \ \zeta^2= 196.212144;\nonumber \\
\epsilon &=& 10^{-4} \RA (3,40,43), \quad \  \zeta^2=
196.212121;\nonumber \eea
The only difference with the case $\epsilon = 0$ is that now the two
co-existing triads are not symmetrical anymore. In Table \ref{t:1}
the distribution of the resonance generating vorticities is given.
One can see immediately that more that $30\%$ of all vorticities are
relatively small, in the sense that they satisfy the condition $
\Omega=2 \, \zeta \,\sqrt{\sigma} \ < 2 \cdot 10^2 \sqrt{\sigma}. \
$ (first row on the left in the Table \ref{t:1}). The number of
resonance generating vorticities in the partial domains is
decreasing exponentially with the growth of the general spectral
domain. Interestingly enough, new asymmetrical triads appear for
$\epsilon$ of the order of $10^{-4}$ or bigger, i.e. $\epsilon$ is
bounded from below  (cf. detuned frequency balance in the
irrotational case \cite{K07,T07}).

 \noindent
\begin{table}[h]
\begin{tabular}{|c|c||c|c|}
  \hline 
  Magnitude of $\tilde{\zeta}$& Number of $\Omega$ & Magnitude of $\tilde{\zeta}$ & Number of $\Omega$  \\ \hline
  $\tilde{\zeta}<1$             & $3477$             & $4 \le \tilde{\zeta} <5$     & $494$               \\ \hline
  $1\le \tilde{\zeta}<2$      & $1146$             & $5 \le \tilde{\zeta} <6$     & $407$               \\ \hline
  $2\le \tilde{\zeta}<3$      & $781$              & $6 \le \tilde{\zeta} <7$     & $363$               \\ \hline
  $3\le \tilde{\zeta}<4$      & $609$              & $7 \le \tilde{\zeta} $    & $2723$               \\ \hline
\end{tabular}
\caption{\label{t:1} The distribution of the magnitudes of the
resonance generating vorticities is presented, in the spectral
domain $k_1, k_2 \le 100.$ The notation
$\tilde{\zeta}=\zeta^2\cdot10^4$ is used.}
\end{table}%

Generically, to describe
 the dynamics of the resonances in the rotational case, two steps
 have to be performed. Firstly,  nonlinear evolution
 equations governing the rotational case should
 be derived. Afterwards the standard procedure would be to construct all resonant
clusters met in the  solution set above and write out explicitly the
corresponding dynamical systems \cite{KM07}. Secondly, the
interaction coefficient
 has to be computed which, of course, will be different from
 the coefficient known for irrotational flows. Computing the explicit form of the interaction
coefficient for a given wave system is a non-trivial technical
problem demanding tedious computations which can not be automatized
in the present state of symbolic programming
\cite{KRFMS}.\\

\noindent \textbf{5. Flows with constant vorticity.}
Let us recall the governing equations for capillary water waves
propagating at the free surface of a layer of water above a flat bed
\cite{Jo}. In the fluid domain of average depth $d>0$ bounded above
by the free surface $z=d+\eta(x,y,t)$ and below by the flat bed
$z=0$, the velocity field ${\bf u}({\bf x},t)$ and the pressure
function $P({\bf x},t)$, where ${\bf x}=(x,y,z)$ and
${\bf u}=(u_1,u_2,u_3)$, satisfy the Euler equation %
\be\label{ge1}
{\bf u}_t+ ({\bf u} \cdot \nabla)\, {\bf u}+\nabla P=0, \ee %
as well
as the equation of mass conservation%
 \be\label{ge2} \nabla \cdot
{\bf u}=0, \ee%
 expressing homogeneity (with constant water density $\rho=1$). These
 equations are coupled with the boundary conditions%
 \be\label{ge3} P=\sigma H,\quad
u_3=\eta_t+u_1 \,\eta_x + u_2\,\eta_y, \ee %
on the free surface
$z=d+\eta(x,y,t)$, the constant $\sigma>0$ being the surface tension
coefficient and $H$ being twice the mean curvature,
$$H(x,y,t)=\frac{(1+\eta_x^2)\eta_{yy}-2\eta_x\eta_y\eta_{xy}+
(1+\eta_y^2)\eta_{xx}}{(1+\eta_x^2+\eta_y^2)^{3/2}}.$$
On the flat bed $z=0$ we require%
 \be\label{ge4} u_3=0. \ee %
 The governing equations
for capillary waves are (\ref{ge1})-(\ref{ge4}). These equations are
well-posed \cite{CSh}: starting with an initial surface profile
$\eta(x,y,0)$ with Sobolev regularity $H^{6.5}$ and an initial
velocity field ${\bf u}({\bf x},0)$ with Sobolev regularity
$H^{5.5}$ satisfying (\ref{ge2}), for some time $t \in [0,T)$ with
$T>0$ there exists a unique solution $({\bf u},\,\eta,\,P)$ of the
system (\ref{ge1})-(\ref{ge4}). In our context it is important to
keep track of the vorticity $\Omega({\bf x}, t)$, obtained at any
time $t$ from the velocity profile as $\nabla \times {\bf
u}=\Omega$. With this purpose, notice that using (\ref{ge1})-(\ref{ge4}) one can
derive the vorticity equation \cite{MB} \be\label{ve} \Omega_t+({\bf
u}\cdot\nabla)\,\Omega=(\Omega\cdot\nabla)\,{\bf u}. \ee
To investigate further the vorticity, it is useful to introduce the flow map ${\bf x} \mapsto
\Phi({\bf x},t)$: this map advances each particle
in the water region from its position ${\bf x}$ at time $t=0$ to its
position $\Phi({\bf x},t)$ at time $t$. For fixed $t$, $\Phi$ is an invertible
smooth mapping and from (\ref{ge1}), (\ref{ge2}), (\ref{ve}) one can infer that \cite{MB}
\be\label{ve2} \Omega(\Phi({\bf x},t),t)=J({\bf x},t)\,
\Omega({\bf x},0), \ee where $J({\bf
x},t)$ is the Jacobian matrix of the flow map. An immediate
consequence of (\ref{ve2}) is that in three-dimensional flows a
particle which has no vorticity never acquires it and conversely, a
particle which is moving rotationally will continue to do so.

\smallskip
\textbf{Theorem 2} (on the dimension of flows with constant vorticity).
{\it Capillary wave trains can propagate at the free surface of a layer of water with a
flat bed in a flow of constant non-zero vorticity only if the flow is two-dimensional.
 } \smallskip

\textbf{Proof.} A wave train is a periodic surface wave which
propagates without change of shape
at constant speed $c>0$ in a fixed direction, say, that of the $x$-coordinate, and which is
unchanged in the $y$-direction (horizontal and orthogonal to the wave propagation
direction). Notice that for a two-dimensional water flow (in our setting, independent of the
$y$-coordinate), we have $\Omega_1=\Omega_3=0$ so that
$(\Omega\cdot\nabla)\,{\bf u}=\Omega_2\, {\bf u}_y=0$ and therefore (\ref{ve}) implies that
$\Omega_t+({\bf u}\cdot\nabla)\,\Omega=0$: the vorticity of
  each individual water particle is conserved as the particle moves about. In particular,
  if initially the vorticity is constant, it will stay so. The existence of wave trains is ensured
  in this setting provided the wave speed is given by the dispersion relation (\ref{dis}) cf. \cite{W2}.

Conversely, consider the wave train $z=d+\eta(x-ct)$ propagating in a flow of
constant vorticity $\Omega \neq 0$.  Then $(\Omega\cdot\nabla)\,{\bf u}=0$ by (\ref{ve}),
  that is, at every instant the vector ${\bf u}$ is constant in the direction of
  $\Omega$. We first claim that for non-flat
  free surfaces, the direction $\Omega$ has to be horizontal. Indeed, if $\Omega_3 \neq 0$, this
  in combination with (\ref{ge4}) would yield that $u_3 \equiv 0$ throughout the flow. The
  third component of (\ref{ge1}) then forces $P_z=0$ so that
  \be\label{p}
  P(x,y,z,t)=\sigma H(x-ct)\ee
  throughout the flow in view of the first equation in (\ref{ge3}). Moreover,
  $\partial_z u_2=-\Omega_1$ and $\partial_z u_1=\Omega_2$ yield
  \be\label{vc}
  \left\{\begin{array}{l}
  u_1(x,y,z,t)=v_1(x,y,t)+\Omega_2 z,\\
  u_2(x,y,z,t)=v_2(x,y,t)-\Omega_1 z.
  \end{array}\right.
  \ee
  Further, from (\ref{ge2}) we infer the existence of a function $\psi(x,y,t)$ satisfying
  \be\label{st}\psi_x=-v_2,\qquad \psi_y=v_1.\ee
 Writing the first two components of (\ref{ge1}) on the flat bed $z=0$ and on $z=\varepsilon>0$ with
 $\varepsilon>0$ small enough for this horizontal plane to be in the fluid domain, we get
 $$\Omega_2 \psi_{xy}-\Omega_1 \psi_{yy}=-\Omega_2 \psi_{xx}+\Omega_1 \psi_{xy}=0.$$
 As $\Omega_3=\partial_x u_2-\partial_y u_1$ yields
 \be\label{la} (\partial_x^2 + \partial_y^2)\psi=-\Omega_3,\ee
 and $(\Omega \cdot \nabla)\,{\bf u}=0$ yields
 $$\Omega_1\psi_{xy}+\Omega_2\psi_{yy}+\Omega_2\Omega_3=-\Omega_1\psi_{xx}-\Omega_2\psi_{xy}-\Omega_1\Omega_3=0,$$
 unless $\Omega_1=\Omega_2=0$ we infer from these relations that
 $$ \psi_{xx} = -\frac{\Omega_1^2\Omega_3}{\Omega_1^2+\Omega_2^2},\
 \psi_{xy} = -\frac{\Omega_1\Omega_2\Omega_3}{\Omega_1^2+\Omega_2^2},\
\psi_{yy} = -\frac{\Omega_2^2\Omega_3}{\Omega_1^2+\Omega_2^2}.$$
Therefore, assuming $\Omega_1^2+ \Omega_2^2>0$, we get
$$\psi(t,x,y)=A\, y^2 + B\, xy + C\, x^2 + a(t)\,x+b(t)\,y +k(t),$$
for some functions $a,\,b,\,k$, where we denoted
$$A= -\frac{\Omega_2^2\Omega_3}{2(\Omega_1^2+\Omega_2^2)},\
B=-\frac{\Omega_1\Omega_2\Omega_3}{\Omega_1^2+\Omega_2^2},\
C = -\frac{\Omega_1^2\Omega_3}{2(\Omega_1^2+\Omega_2^2)}.$$
From (\ref{ge1}) we now infer that
$$P_x= -b'(t)-B\,b(t)+2A\,a(t),\ P_y = a'(t)+2C\,b(t)-B\,a(t),$$
and (\ref{p}) then yields
$$P(x,y,t)=\alpha\,(x-ct)+\beta$$
for some constants $\alpha,\,\beta$. But then $P=\sigma H$ is impossible since the
right-hand side has by periodicity infinitely many zeros (at least one between 
two consecutive crests, by the mean-value theorem). To rule out the possibility that
$\Omega_3 \neq 0$, it remains to
consider the case $\Omega_1=\Omega_2=0$. If this holds, then
the functions $u_1$ and $u_2$ are independent of $z$ in view of (\ref{vc}) and are
harmonic in $(x,y)$ by (\ref{st})-(\ref{la}). The second equation
  in (\ref{ge3}) yields $[u_1(x,y,t)-c]\,\eta_x(x-ct)=0$ for all $x,\,y$ real. But the function $u_1$
  is a harmonic function of $(x,y)$ and taking into account the structure of the
  level sets of a harmonic function (they
  are curves in the plane unless the function is constant cf. \cite{C}), this yields $u_1 \equiv c$
  since $\eta_x \not \equiv 0$. But then the first equation in (\ref{ge1}) ensures $P_x=0$, while
  (\ref{ge2}) yields that $u_2$ is independent of $y$. As $\Omega_3=\partial_x u_2-\partial_y u_1$
  we have $u_2(t,x,y)=\Omega_3 x+f(t)$ for some function $f$. Now the second
  equation in (\ref{ge1}) yields $P_y=-f'(t)-c\Omega_3$. But this can be reconciled with $P(x,y,t)=\sigma H(x-ct)$
  only if $H$ is a constant and then the surface has to be flat \cite{A, M}.

  We therefore proved that $\Omega_3=0$. It remains to show that $\Omega_1=0$ and that
  $P$ and ${\bf u}$ are independent of $y$. Since
  $\Omega_1 \partial_xu_i + \Omega_2 \partial_y u_i=0$ for $i =1,2,3$ and $\nabla \times
  {\bf u}=\Omega$ throughout the flow, we first infer that
  \be\label{u2}\Omega_1  u_1 + \Omega_2 u_2=h(t)\ee
  for some function $h$, since all spatial derivatives of the left-hand side are zero. Multiplying
  the first equation in (\ref{ge1}) by $\Omega_1$, the second equation by $\Omega_2$, and
  adding up, we infer from the above relations that
  \be\label{pe1}\Omega_1 P_x+\Omega_2 P_y=-h'(t)\ee
  throughout the fluid.  Notice that once we show that $\Omega_1=0$,
  then $\Omega_2 \neq 0$ as $\Omega \neq 0$, and $(\Omega \cdot \nabla){\bf u}=0$
  ensures that ${\bf u}$ does not depend on $y$.
  Moreover, in this case (\ref{u2}) yields that $u_2$ is simply a function of time and consequently the second
  equation in (\ref{ge1}) forces $P$ to be independent of $y$. To complete the proof it therefore suffices to
  show that $\Omega_1=0$. Assuming $\Omega_1 \neq 0$, the relation $\partial_x u_2-\partial_y u_1=0$
  ensures the existence of a function $\varphi({\bf x},t)$ satisfying
  $$u_1=\varphi_x,\quad u_2=\varphi_y.$$
  From $\nabla \times {\bf u}=\Omega$ we infer
  $$\partial_y u_3=\varphi_{yz}+\Omega_1,\quad \partial_x u_3=\varphi_{xz}-\Omega_2,$$
  so that, adding if necessary a function depending only on $(z,t)$ to $\varphi$, we have
  $$u_3=\varphi_z + \Omega_1 y - \Omega_2 x.$$
  Denoting $\xi=\Omega_2 x - \Omega_1 y$, we get
  \be\label{us}
  u_1=\varphi_x,\quad u_2=\varphi_y,\quad u_3=\varphi_z - \xi,\ee
  and (\ref{ge2}) becomes
  \be\label{us2}
  \varphi_{xx}+\varphi_{yy}+\varphi_{zz}=0.\ee
  From (\ref{u2}) and (\ref{pe1}) we deduce that $\varphi$ and $P$ admit the decomposition
  \be\label{us3} \left\{\begin{array}{ccc}
  \varphi({\bf x},t) &=& (\mu x + \nu y)\,h(t)+F(\xi,z,t),\\
  P({\bf x},t) &=& - (\mu x+ \nu y)\,h'(t) + G(\xi,z,t),
  \end{array}\right.\ee
  for some functions $F,\,G$, where we denoted
  $$\mu=\frac{\Omega_1}{\Omega_1^2+\Omega_2^2},\ \nu=\frac{\Omega_2}{\Omega_1^2+\Omega_2^2}.$$
  Using (\ref{us}) and (\ref{us3}), we can express (\ref{ge1}) as
  $$\left\{\begin{array}{l}
  \Omega_2 \,\partial_\xi \,[ F_t+\frac{\Omega_1^2+\Omega_2^2}{2}\,F_\xi^2 +
  \frac{1}{2}\,F_z^2 - \xi F_z +G]=0,\\[0.1cm]
  \Omega_1 \,\partial_\xi \,[ F_t+\frac{\Omega_1^2+\Omega_2^2}{2}\,F_\xi^2 +
  \frac{1}{2}\,F_z^2 - \xi F_z +G]=0,\\[0.1cm]
  \partial_z \,[ F_t+\frac{\Omega_1^2+\Omega_2^2}{2}\,F_\xi^2 +
  \frac{1}{2}\,F_z^2 - \xi F_z +G]=(\Omega_1^2+\Omega_2^2)\,F_\xi.
  \end{array}\right.$$
  Under our working assumption $\Omega_1 \neq 0$ these relations imply $F_{\xi\xi}=0$. But
  since (\ref{us2}) and (\ref{us3}) yield
  $$F_{zz}+(\Omega_1^2+\Omega_2^2)\,F_{\xi\xi}=0,$$
  we also have $F_{zz}=0$. Consequently
  $$F(\xi,z,t)=f_0(t)\,\xi z + f_1(t)\,\xi+f_2(t)\,z + f_3(t)$$
  for some functions $f_i$, $i=0,1,2,3$. From (\ref{us}) and (\ref{us3}) we now infer that
  $$\left\{\begin{array}{ccc}
  u_1({\bf x},t) &=& -\,\Omega_2 \,[f_0(t)\,z+f_1(t)]+\mu\,h(t),\\
  u_2({\bf x},t) &=& \Omega_1\,[f_0(t)\,z+f_1(t)]+\nu\,h(t),\\
  u_3({\bf x},t) &=& [f_0(t)-1]\,( \Omega_2 x - \Omega_1 y)+f_2(t).
  \end{array}\right.$$
  But then equating the coefficient of $y$ on
  both sides of the second relation in (\ref{ge3}), stating that
  \be\label{fe}u_3=(u_1-c)\,\eta_x(x-ct)\quad\hbox{on}\quad z=d+\eta(x-ct),\ee
  we first infer that $f_0(t) \equiv 1$ as $\Omega_1 \neq 0$ by assumption. But then $u_3({\bf x},t)=f_2(t)$ and 
  now the right-hand side of (\ref{fe}) is periodic in $x$ and
  the left-hand side is independent of $x$. Therefore $\eta_x \equiv 0$ which contradicts the fact that the free 
  surface was not flat. This concludes the proof.\smallskip

 Theorem 2 allows us to consider for flows of constant vorticity only two-dimensional flows propagating
  in the $x$-direction, for which the vorticity vector takes the form $(0,\Omega_2,0)$. For such flows it is
  therefore natural by abuse of notation to identify the vorticity vector $\Omega$ with its second component. For irrotational
  flows $\Omega=0$ so that a velocity potential exists and
methods from harmonic function theory (involving the
Dirichlet-Neumann operator) can be used to transform the governing
equations for capillary water waves to a Hamiltonian system
expressed solely in terms of the free surface $\eta(x,t)$ and of the
restriction of the velocity potential on the free surface \cite{Z}.
The absence of a velocity potential for non-zero
vorticities complicates the analysis considerably cf. \cite{CS2}.\\

\noindent \textbf{6. Discussion.} Our main conclusions can be
formulated as follows:

\textbullet~ Flows with constant non-zero vorticity admit capillary
wave trains only if they are two-dimensional.

\textbullet~ Only positive vorticity  can trigger the appearance of
three-wave resonances.

\textbullet~ The number of positive vorticities which do trigger a
resonance is countable.

\textbullet~ The magnitude of a positive vorticity triggering a resonance can not be too small.

It is remarkable that relatively large positive constant vorticities
are necessary to observe resonant 3-wave interactions. The fact that
resonances can never occur for negative constant vorticities
substantiates the belief that the vorticity of the flow has a big
influence on the dynamics of the surface water waves. In this
context the ``frozen turbulence" observed in numerical simulations
\cite{PZ00} for capillary waves is just a manifestation of the fact
that resonances are absent.

To understand the nonlinear resonance dynamics among rotational
capillary waves,  an evolution  equation corresponding to the
rotational case should be derived. The
 presence of non-zero vorticity
invalidates the existence of a velocity potential  for the
flow, as is the case for irrotational flows, and harmonic function
theory is not readily available for the analysis. Two common methods
of analysis to find the exact form of the dynamic equations are the
method of multiple scales and variational techniques \cite{HH}. Both
methods are applicable for flows with vorticity \cite{Jo, CSS, CS2, W2}; this
is work in progress.
\\

\noindent \textbf{Acknowledgements.} E.K. acknowledges
the support of the Austrian Science Foundation (FWF) under the project
P20164-N18 ``Discrete resonances in nonlinear wave systems".


\end{document}